\documentclass[a4paper,11pt]{article}
\pdfoutput=1
\usepackage{jinstpub} 
\usepackage{lineno}
\title{\boldmath Design, construction, characterization, and testing of one-channel electronic module for the Hamamatsu Multi-Pixel Photon Counter S12572-100P}

 \author{L. Arceo,}
 \author{ J. F\'{e}lix}
 \affiliation{Laboratorio de Part\'{i}culas Elementales, Divisi\'{o}n de Ciencias e Ingenier\'{i}as campus Le\'{o}n, Universidad de Guanajuato,\\Loma del Bosque, 103, Col. Lomas del Campestre, C.P. 37150, Le\'{o}n, Guanajuato, M\'{e}xico}

\emailAdd{miquel@fisica.ugto.mx}
\abstract{The Multi-Pixel Photon Counter (MPPC) or Silicon Photomultiplier (SiPM) can detect from a single photon to several thousand ones; it has high gain ---$10^{5}$ to $10^{6}$---, low operation voltage ---65 to 75 Vdc---, and small size ---typically 3.85 mm $\times$ 4.35 mm $\times$ 1.45 mm---; it is suitable for very low light level detection applications, like in cosmic ray detectors, in high-energy particle detectors, and in ionizing radiation detection in general. To study, test, and apply the Hamamatsu Multi-Pixel Photon Counter S12572-100P, we proposed a three-electronic-board MPPC module to connect, to feed and read-out, and to digitize its signal. The MPPC module waveband operation is from 100 Hz to 1 MHz; the ratio of outgoing signal to incoming signal (transmission coefficient) is about 82 \% between 100 Hz--10 kHz and 72 \% between 10 kHz--1 MHz; the phase shifts between the incoming and outgoing signals were close to zero degrees for all frequencies and occasionally, different from zero degrees; signal transit time is about (741.72$\pm$82.80) ps; the digitizing efficiency error is 0.88$\times10^{-3}$ \%; the digitizing efficiency is 99.99 \%; digitizing time is about (2.96$\pm$0.13) ns. We report on technical details and physical characteristics of this electronic module to drive the Hamamatsu Multi-Pixel Photon Counter S12572-100P, and on preliminary physical results obtained applying this electronic module.}


\begin{document}
\maketitle
\flushbottom

\section{Introduction}
MPPCs offer advantages over traditional Photomultiplier Tubes (PMTs). In many fields of physics, the applications are many, like data communications, aerospace, laser range finders, radiation detection in high energy physics and cosmic ray physics, medicine, and other areas of science, due to their physical qualities like high speed, high sensitivity, high quantum efficiency, short rising time, broad wavelength range of detection, unaffected by external magnetic fields, good operation at moderate room temperature, small volume ---typically 3.85 mm $\times$ 4.35 mm $\times$ 1.45 mm---, small operation voltage, significant amplification factor, and low-cost.

The MPPC made a high-density matrix of Single-Photon Avalanche Photodiode (SPAD) pixels operating in Geiger mode; these pixels are arranged in two dimensions and have high internal gain, which enables single-photon detection to a lot of photon detection. Its advantages, such as high internal gain and low operation voltage, make the MPPC a good option for light detection from a single photon to thousand photons ~\cite{A-Hamamatsu1,B-Gundacker_2020} in particular in applications with very low light levels; this last feature is perfect for developing cosmic ray detectors, where radiation level is very low ~\cite{C-Hamamatsu2,D-Hamamatsu3,E-YOKOYAMA2009128}.

Between all MPPCs, the Hamamatsu Multi-Pixel Photon Counter S12572-100P is an excellent option, due to its good physical properties and low-cost, on many applications previously explained.

Normally, the Hamamatsu company only supplies a circuit hint on how to connect and feed this device, leaving the user with many open possibilities on how to use it.

To use the Hamamatsu Multi-Pixel Photon Counter S12572-100P, we designed, constructed, characterized, and tested a three-electronic-board MPPC module, to connect, to feed and read out, and digitize its signal, with the following characteristics: 1) it is designed for the Hamamatsu Multi-Pixel Photon Counter S12572-100P to be attached to the detection material and connected to this module; 2) it has response time in the order of nanoseconds; 3) it is adaptable to another Hamamatsu MPPC series. For example, the Hamamatsu optical measurement module C15524-1310SA, that has a MPPC with similar characteristics, but does not digitize; for this reason, we developed our own MPPC module. The MPPC module was tested with a CompactRIO (cRIO) Data Acquisition System (DAQ) ~\cite{F-UserManual-cRIO-9025,G-UserManual-cRIO-9118,H-Datasheet-cRIO-9402} from National Instruments.

In the following sections we present technical details of the MPPC module: design; construction; characterization and operations; we explain the experimental setup; describe and discuss results, and give conclusions.

\section{Description of the MPPC module}
In this section, we provide technical description of the design, construction, operation and characterization of three-electronic-board MPPC module.

The MPPC module has three electronic boards:

\begin{enumerate}
\item Connection electronic board, where this Hamamatsu Multi-Pixel Photon Counter S12572-100P is placed to attach it mechanically to the detector material.
\item Feeding and reading-out electronic board, to apply the feeding voltage, create the analogical signal, and read it out.
\item Digitizing-electronic board, where the analogical signal is digitized and sent out to the oscilloscope or DAQ system.
\end{enumerate}

All of them were separately planned, designed, manufactured and tested. We report on the transmission coefficient of the analogical signal as function of frequency, the phase shift outgoing signal with respect to ingoing signal, transit time, digitizing efficiency error, digitizing efficiency, and digitizing time.

\subsection{Connection electronic board}
\begin{figure}[ht!]
\centering % \begin{center}/\end{center} takes some additional vertical space
\includegraphics[width=0.98\textwidth,origin=c]{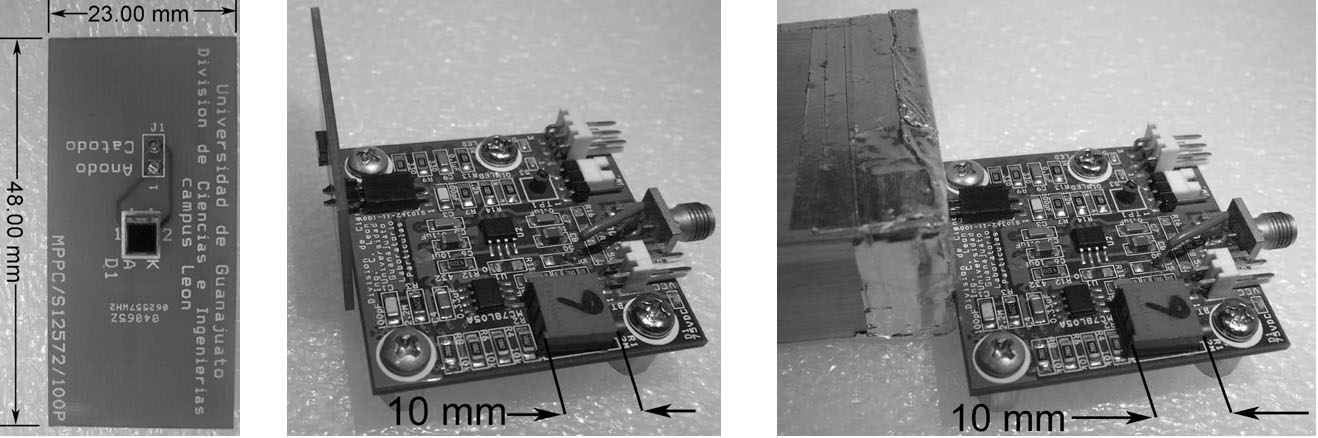}
\caption{\label{Fig:ConElecBoard} Left: Connection electronic board, top layer side. Middle: Connection electronic board connected to the feeding and reading-out electronic board. Right: Connection electronic board attached and assembled to the detection material and connected to the feeding and reading-out electronic board.}
\end{figure}

The electronic board physically separates the Hamamatsu Multi-Pixel Photon Counter S12572-100P from its electronics and attaches it mechanically to the detection material (an 1-in $\times$ 2-in $\times$ 8-in solid aluminum bar, with one 1-in $\times$ 2-in end polished to mirror-like finishing). In figure~\ref{Fig:ConElecBoard} (left) we show the manufactured connection electronic board. It is where the Hamamatsu Multi-Pixel Photon Counter S12572-100P surface mount type is soldered on the top layer. In figure~\ref{Fig:ConElecBoard} (middle) we show the connection electronic board connected to the feeding and reading-out electronic board without detection material. In figure~\ref{Fig:ConElecBoard} (right) we show the connection electronic board attached and assembled to the detection material and connected to the feeding and reading-out electronic board.

\subsection{Feeding and reading-out electronic board}
\begin{figure}[ht!]
\centering
\includegraphics[width=.95\textwidth,trim=0 0 0 0,clip]{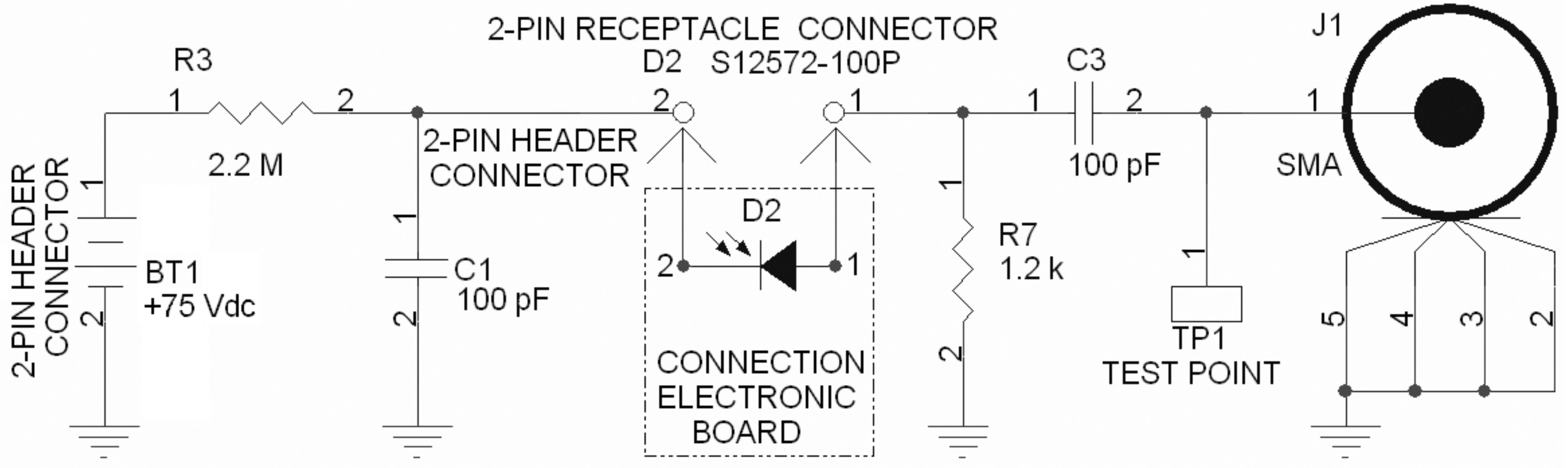}
\caption{\label{Fig:FRElecBoard}
Schematic diagram design of the feeding and reading-out electronic board for the Hamamatsu Multi-Pixel Photon Counter S12572-100P.}
\end{figure}

The electronic board was designed, constructed, and operated to feed the Hamamatsu Multi-Pixel Photon Counter S12572-100P, and to read-out its analog signal. In figure~\ref{Fig:FRElecBoard} and figure~\ref{Fig:ConElecBoard} (middle), we show the schematic diagram design and manufactured electronic board of the feeding and reading-out electronic board, respectively.

%%Transmission coefficient and phase shift
In this electronic board, we measured three parameters:
\begin{enumerate}
\item The transmission coefficient ---outgoing signal amplitude divided by ingoing signal amplitude --- as function of frequency.

\item The phase shift, which is the phase displacement of the output signal waveform with respect to the input signal.

To measure these parameters, we replaced the Hamamatsu Multi-Pixel Photon Counter S12572-100P analog signal by a Tektronix function generator AFG3101 ranging from 100 Hz to 1 MHz. To measure the input amplitude, the output amplitude, and the phase shift between signals, the Tektronix oscilloscope TDS2022C-EDU was employed; from the oscilloscope, we consider a systematic error of 10 \% ~\cite{I-UserManual-TDS2022C-EDU,J-use_of_an_oscilloscope}; it was included in the least square technique to measure the transmission coefficient as a function of the frequency. To record the amplitudes and phase shift of the signals, a PC-host with LabVIEW\textsuperscript{TM} was used.

\item The transit time, or propagation delay, or response time ~\cite{K-coughlin2001operational}, ---the time an electrical signal takes to travel through the MPPC module---. 

To measure this parameter, we used a Tektronix function generator AFG3101 at 1 MHz, and a Tektronix oscilloscope TDS5104B, to measure the delay between input and output signals.
\end{enumerate}

\subsection{Digitizing-electronic board}
\label{DigElecBoard}
\begin{figure}[ht!]
\centering
\includegraphics[width=1.00\textwidth,trim=0 0 0 0,clip]{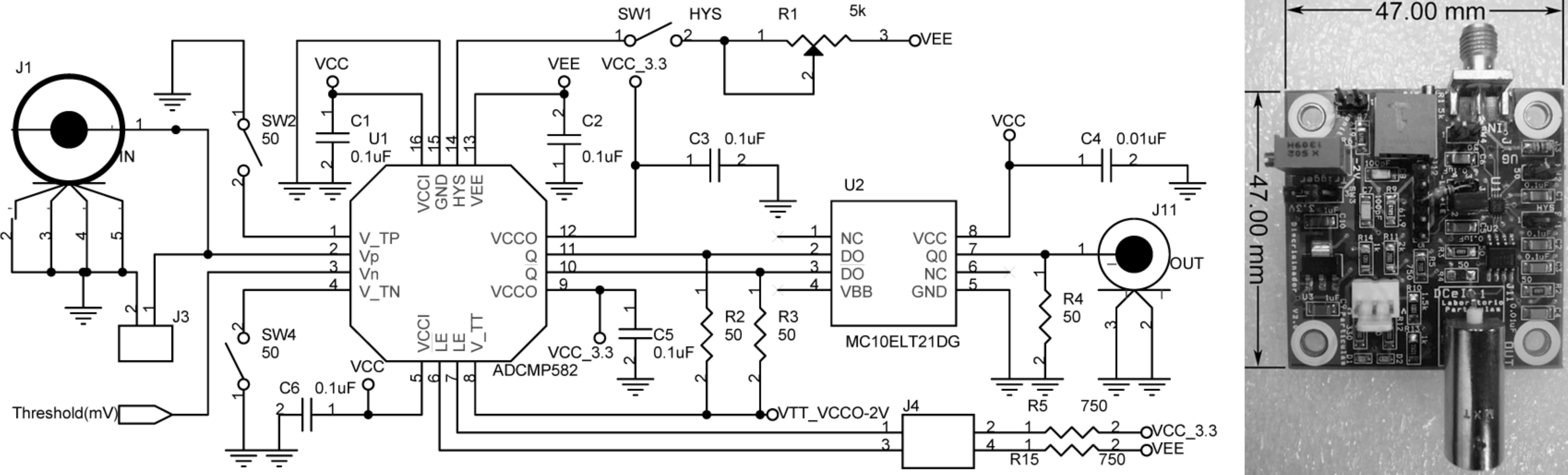}
\caption{\label{Fig:SchDiagDigitElecB}
Left: Schematic diagram of the digitizing-electronic board; right: Manufactured digitizing-electronic board.}
\end{figure}

This electronic board was designed to compare, against the user choice threshold voltage (three times above the noise signal), and 1-bit digitize the feeding and reading-out electronic board analog signal. Figure~\ref{Fig:SchDiagDigitElecB} (left) and (right) show the schematic diagram design and manufacture of the digitizing-electronic board, respectively.

The ADCMP582 chip (U1), a PECL (positive emitter-coupled logic) output driver, has a typical one-channel propagation delay of 180 ps ~\cite{L-Datasheet-ADCMP582}; and MC10ELT21DG chip (U2), PECL to LVTTL family conversion (for achieving cRIO compatibility), has a typical propagation delay of 3.5 ns  ~\cite{M-Datasheet-MC10ELT21DG}.

%%Digitizing efficiency and digitizing efficiency error
In this electronic board, we measured these three parameters:
\begin{enumerate}
\item The digitizing efficiency error $eff$ (\%), which is defined by the following equation,

\begin{equation}\label{eq:eff}
%eff=\frac{1}{n}\sum_{j=1}^{n}\frac{f_{analog}^{in}-f_{j,digital}^{out}}{f_{analog}^{in}} \times 100,
eff=\frac{1}{n}\sum_{j=1}^{n}\frac{\omega_{j}-\omega}{\omega} \times 100,
\end{equation}

Here $\omega$ is the frequency of the incoming generated function, with frequency between 1 Hz to 1 MHz with base-10 logarithmic increments. $\omega_{j}$ is the measured frequency of the outgoing measured function, j is an iteration number, n is the total number of measurements ---1800, one measurement per second---.

\item The digitizing efficiency (\%) is obtained by subtracting the digitizing efficiency error (\%) from 100 percent; it depends on the frequency.

\item The digitizing time, or conversion time, ---the time taken by the electronic board to discriminate and convert the analog signal into a digital one---; it is measured using a Tektronix function generator AFG3101 at 1 MHz and a Tektronix oscilloscope TDS5104B.
\end{enumerate}

\section{Experimental setup}
In figure~\ref{Fig:BlockDiaExpSet}, we show a block diagram of the experimental setup; we run two tests: 

\begin{figure}[ht!]
\centering
\includegraphics[width=.98\textwidth,trim=0 0 0 0,clip]{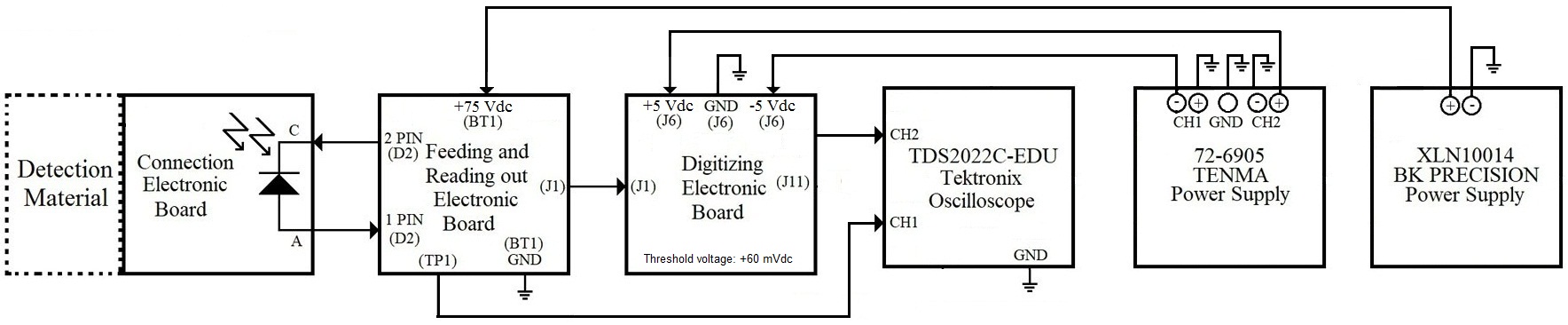}
\caption{\label{Fig:BlockDiaExpSet}
Block diagram of the complete experimental setup, for testing it with green blinking LED and natural radiation.}
\end{figure}

\begin{enumerate}
\item A green LED light was attached to the photosensitive area of 3-mm $\times$ 3-mm of the Hamamatsu Multi-Pixel Photon Counter S12572-100P, and operated inside a black box as follows: 1) with the LED blinking at frequency 1Hz--100kHz, with exponential decay waveform or pulse waveform, and 2) with the LED turned off. 

\item The detection material was attached, by its polished end, to the photosensitive area of 3-mm $\times$ 3-mm of the Hamamatsu Multi-Pixel Photon Counter S12572-100P, outside the black box, and exposed to natural radiation, see figure~\ref{Fig:BlockDiaExpSet}. 
\end{enumerate}

\section{Results from feeding and reading-out electronic board characterization}
We present the results from the characterization of the electronic board. 

\subsection{Transmission coefficient}
\begin{figure}[ht!]
\centering % \begin{center}/\end{center} takes some additional vertical space
\includegraphics[width=.47\textwidth,origin=c]{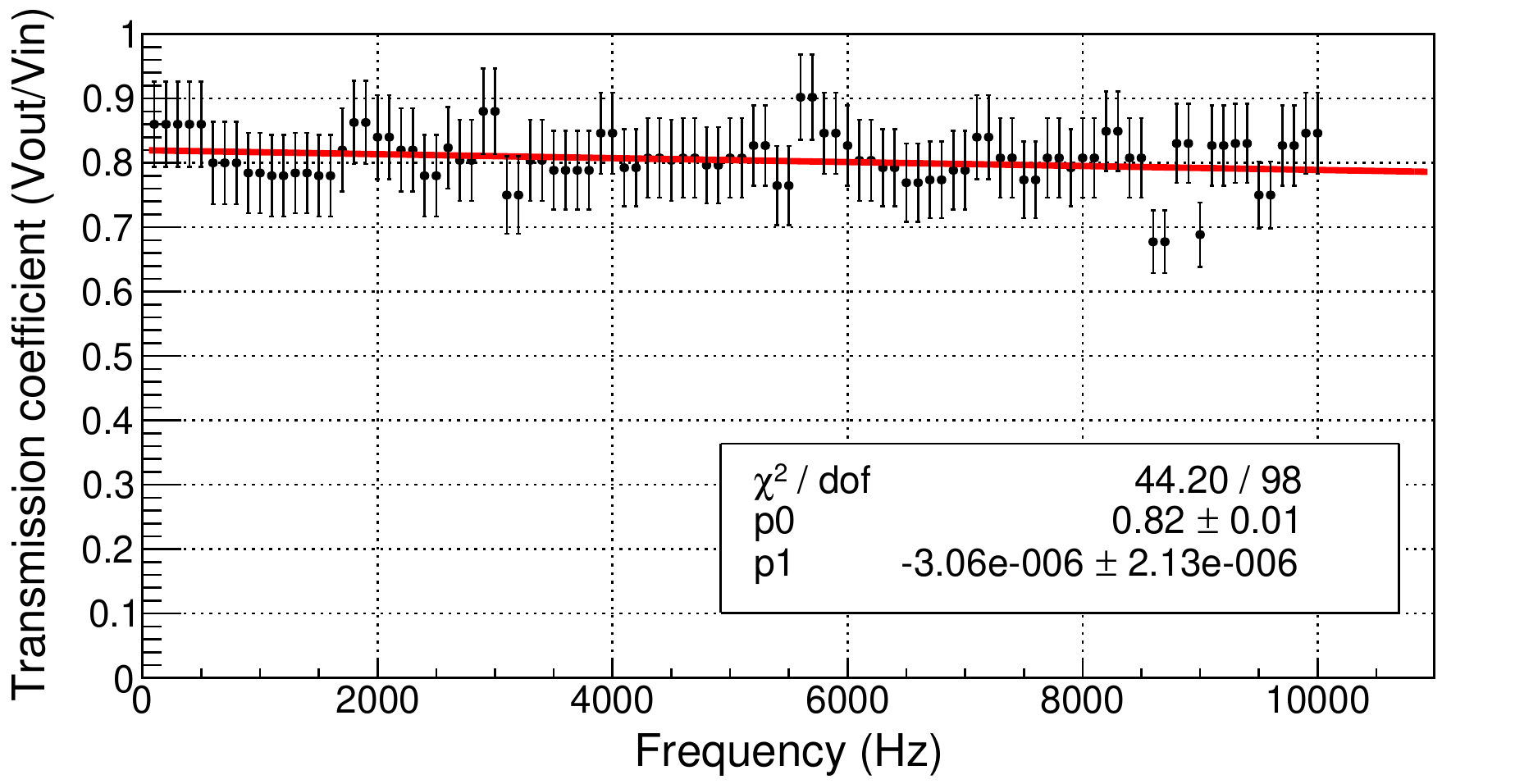}
\qquad
\includegraphics[width=.47\textwidth,origin=c]{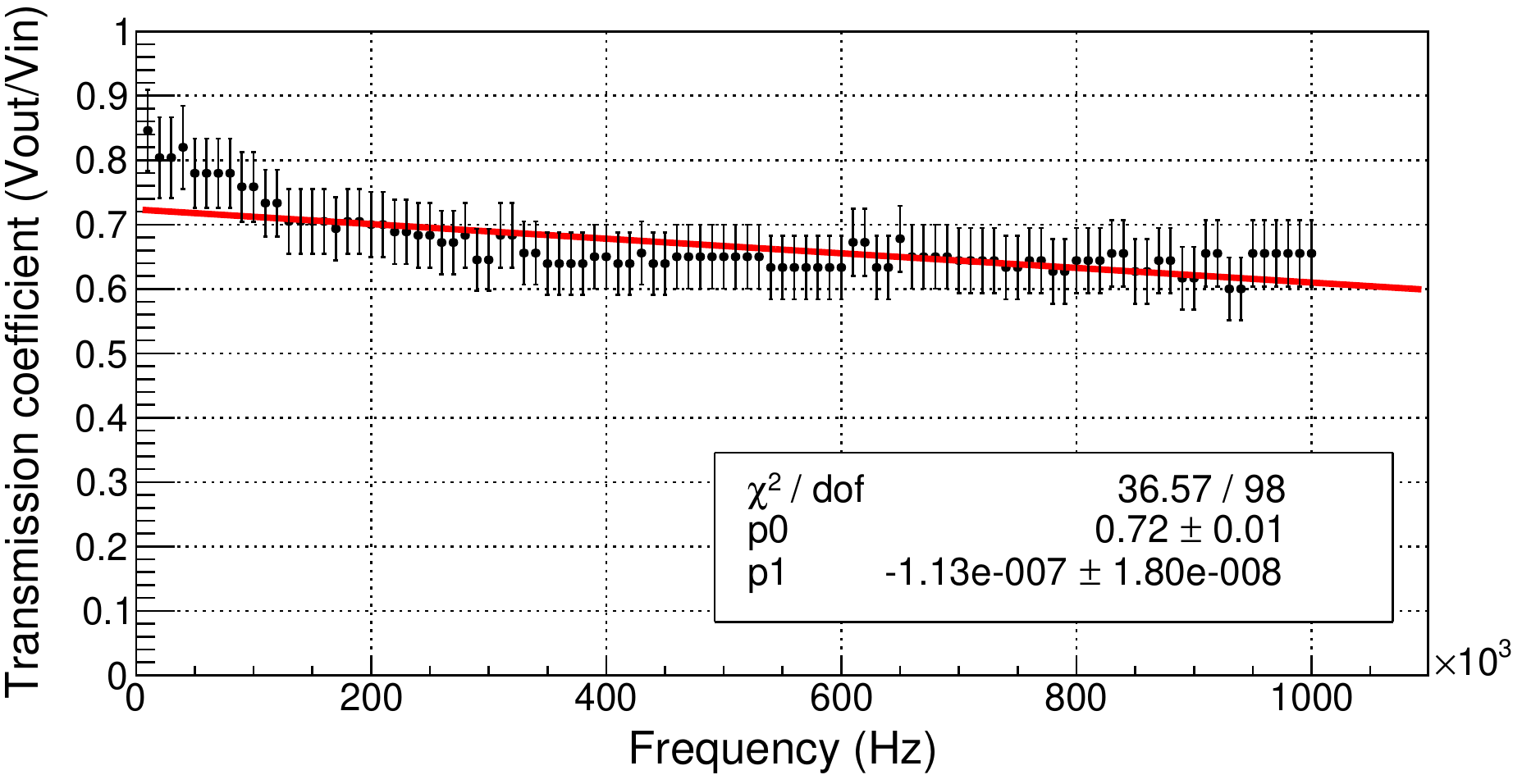}
\caption{\label{Fig:AT} Left: transmission coefficient as a function of the frequency between 100 Hz to 10 kHz; right: transmission coefficient as a function of the frequency between 10 kHz to 1 MHz; the two figures correspond to feeding and reading-out electronic board; a linear fit is superimposed.}
\end{figure}

In figure~\ref{Fig:AT}, we show the results of the transmission coefficient of the feeding and reading-out electronic board as a function of the injected signal frequency.

 On the left side, we plot from 100 Hz to 10 kHz; on the right, from 10 kHz to 1 MHz. We include the linear fit parameters and the $\chi^{2}$ per dof (the number of degrees of freedom), p0 (the Y-axis intercept, which is the average transmission coefficient of the feeding and reading-out electronic board), and p1 (the slope). The p0 values are 0.82$\pm$0.01 and 0.72$\pm$0.01 for 100 Hz to 10 kHz and 10 kHz to 1 MHz, respectively; on average, it is 0.77$\pm$0.01; the p1's are near zero. These results are used for transmission correction and calibration.

\subsection{Phase shift}
\begin{figure}[ht!]
\centering
%\qquad
\includegraphics[width=.47\textwidth,origin=c]{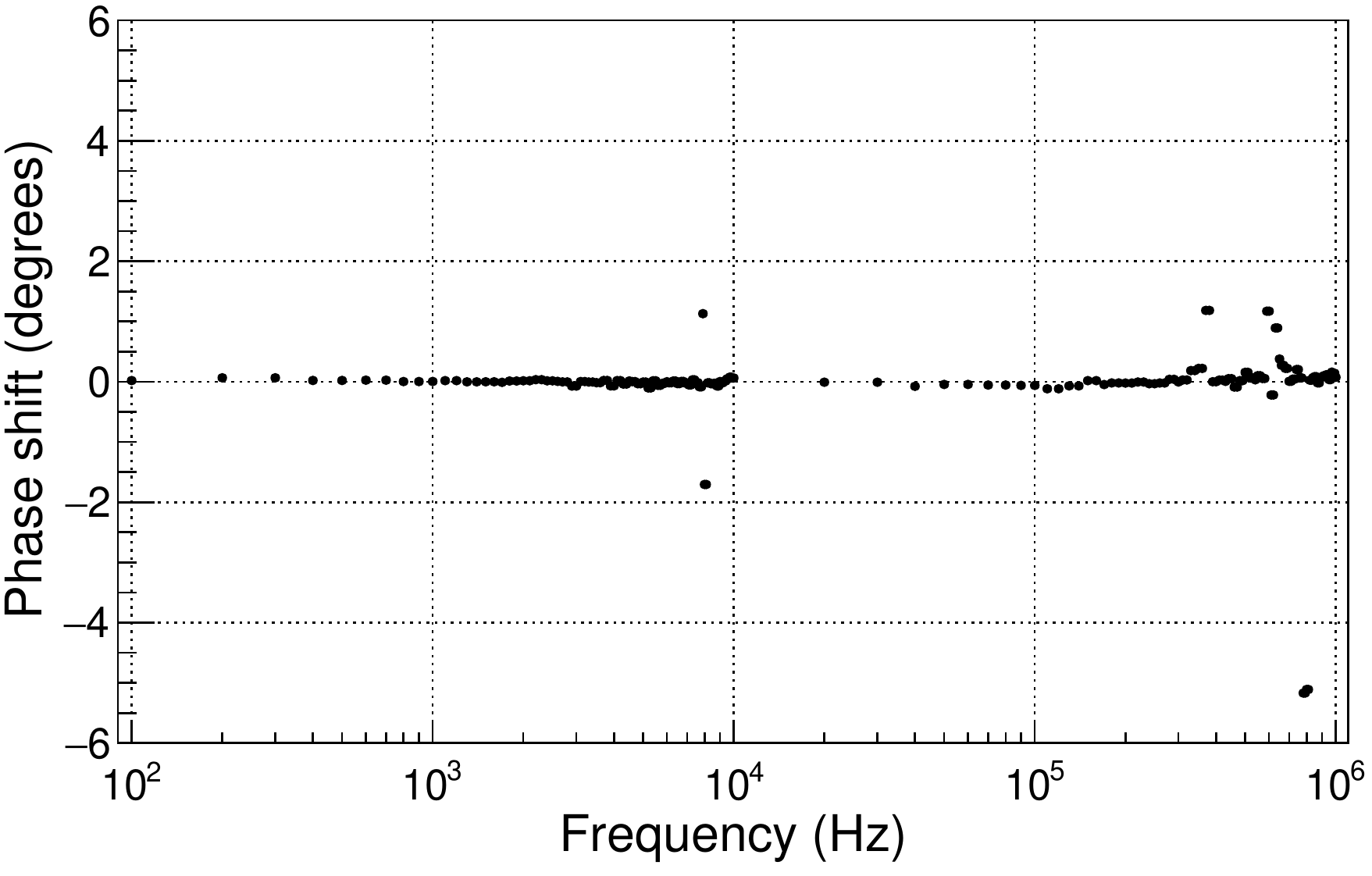}
    \caption{\label{Fig:PS199m} Phase shift measurements vs frequency for 100 Hz to 1 MHz from feeding and reading-out electronic board.}
\end{figure}

In figure~\ref{Fig:PS199m}, we show the results of the phase shift measurements on the feeding and reading-out electronic board as function of the frequency of the injected signal.

We plot from 100 Hz to 1 MHz (199 measurements); at frequencies near to 8 kHz, there are two values for the phase shift less than two degrees; at 370 kHz and 600 kHz, there are four values for the phase shift higher than one degree and less than two degrees, not distinguished; at 800 kHz, there are four values for the phase shift higher than four degrees and less than six degrees, not distinguished. Usually, the phase shift is almost zero degrees and not depending on the frequency; from 199 measurements, only ten were found with values greater than one degree and less than six degrees at frequencies of 10 kHz, 370 kHz, and 600 kHz. These results do no compromise the function of the feeding and reading-out electronic board. Signals are not filtered.

\subsection{Transit time}
The measured transit time was (741.72$\pm$82.80) ps. This value indicates high-speed electronic board, which can read out high-speed signals and avoid superposition of signals.

\section{Results from digitizing-electronic-board characterization}
We present the results from the characterization of the electronic board.

\subsection{Digitizing efficiency error}
In table~\ref{tab:DigitizingEfficiencyError}, we show the results of the digitizing efficiency error ---using eq.~\eqref{eq:eff}--- of the digitizing-electronic board as function of the input frequency. At low frequencies (1 Hz up to 10 Hz), the $eff$ is zero for the electronic board and the function generator. At intermediate frequencies (100 Hz up to 1 MHz), it is slightly increased. Similar dependence is observed at higher frequencies (100 kHz up to 1 MHz).

\begin{table}[htbp]
\centering
\caption{\label{tab:DigitizingEfficiencyError} Digitizing efficiency error (\%) for the digitizing-electronic board and for the Tektronix function generator AFG3101, as function of frequency.}
\begin{tabular}{c|cccccccc}
\hline\hline                      
\begin{tabular}[c]{@{}c@{}}Frequency\end{tabular} & 
\begin{tabular}[c]{@{}c@{}}1 Hz\end{tabular} & 
\begin{tabular}[c]{@{}c@{}}10 Hz\end{tabular} &
\begin{tabular}[c]{@{}c@{}}100 Hz\end{tabular} &
\begin{tabular}[c]{@{}c@{}}1 kHz\end{tabular} &
\begin{tabular}[c]{@{}c@{}}10 kHz\end{tabular} &
\begin{tabular}[c]{@{}c@{}}100 kHz\end{tabular} &
\begin{tabular}[c]{@{}c@{}}1 MHz\end{tabular} & \\
\hline	  
\begin{tabular}[c]{@{}c@{}}Electronic board \\$eff$ ($\times10^{-3}$ \%)\end{tabular} & 
\begin{tabular}[c]{@{}c@{}}0.00\end{tabular} & 
\begin{tabular}[c]{@{}c@{}}0.00\end{tabular} &
\begin{tabular}[c]{@{}c@{}}1.11\end{tabular} &
\begin{tabular}[c]{@{}c@{}}1.28\end{tabular} &
\begin{tabular}[c]{@{}c@{}}1.27\end{tabular} &
\begin{tabular}[c]{@{}c@{}}1.26\end{tabular} &
\begin{tabular}[c]{@{}c@{}}1.24\end{tabular} & \\
\hline	  
\begin{tabular}[c]{@{}c@{}}Function generator\\$eff$ ($\times10^{-3}$ \%)\end{tabular} & 
\begin{tabular}[c]{@{}c@{}}0.00\end{tabular} & 
\begin{tabular}[c]{@{}c@{}}0.00\end{tabular} &
\begin{tabular}[c]{@{}c@{}}1.11\end{tabular} &
\begin{tabular}[c]{@{}c@{}}1.11\end{tabular} &
\begin{tabular}[c]{@{}c@{}}1.12\end{tabular} &
\begin{tabular}[c]{@{}c@{}}1.14\end{tabular} &
\begin{tabular}[c]{@{}c@{}}1.12\end{tabular} & \\
\hline
\hline                     
\end{tabular}
\end{table}

The average $eff$ was 0.88$\times10^{-3}$ \%; for this work, it is negligible.

\subsection{Digitizing efficiency}

In table~\ref{tab:DigitizingEfficiency}, we show the results of the digitizing efficiency ---using eq.~\eqref{eq:eff} and subtracting it from 100 \%--- of the digitizing-electronic board as function of the input frequency. At 1 Hz to 10 Hz, the digitizing efficiency is 100 \% for the electronic board and the function generator. It is slightly decremented (0.01 \%) at frequencies 100 Hz to 1 MHz. 

%\begin{table}[ht!]
\begin{table}[htbp]
\centering
\caption{\label{tab:DigitizingEfficiency} Digitizing efficiency (\%) for the digitizing-electronic board and the Tektronix function generator AFG3101, as function of incoming signal frequency.}
\begin{tabular}{c|cccccccc}
\hline\hline                      
\begin{tabular}[c]{@{}c@{}}Frequency\end{tabular} & 
\begin{tabular}[c]{@{}c@{}}1 Hz\end{tabular} & 
\begin{tabular}[c]{@{}c@{}}10 Hz\end{tabular} &
\begin{tabular}[c]{@{}c@{}}100 Hz\end{tabular} &
\begin{tabular}[c]{@{}c@{}}1 kHz\end{tabular} &
\begin{tabular}[c]{@{}c@{}}10 kHz\end{tabular} &
\begin{tabular}[c]{@{}c@{}}100 kHz\end{tabular} &
\begin{tabular}[c]{@{}c@{}}1 MHz\end{tabular} & \\
\hline	  
\begin{tabular}[c]{@{}c@{}}Electronic\\ board \end{tabular} & 
\begin{tabular}[c]{@{}c@{}}100.00\end{tabular} & 
\begin{tabular}[c]{@{}c@{}}100.00\end{tabular} &
\begin{tabular}[c]{@{}c@{}}99.99\end{tabular} &
\begin{tabular}[c]{@{}c@{}}99.99\end{tabular} &
\begin{tabular}[c]{@{}c@{}}99.99\end{tabular} &
\begin{tabular}[c]{@{}c@{}}99.99\end{tabular} &
\begin{tabular}[c]{@{}c@{}}99.99\end{tabular} & \\
\hline 
\begin{tabular}[c]{@{}c@{}}Function\\ generator\end{tabular} & 
\begin{tabular}[c]{@{}c@{}}100.00\end{tabular} & 
\begin{tabular}[c]{@{}c@{}}100.00\end{tabular} &
\begin{tabular}[c]{@{}c@{}}99.99\end{tabular} &
\begin{tabular}[c]{@{}c@{}}99.99\end{tabular} &
\begin{tabular}[c]{@{}c@{}}99.99\end{tabular} &
\begin{tabular}[c]{@{}c@{}}99.99\end{tabular} &
\begin{tabular}[c]{@{}c@{}}99.99\end{tabular} & \\
\hline
\hline                     
\end{tabular}
\end{table} 

The average digitizing efficiency was 99.99 \%.

\subsection{Digitizing time}
The digitizing time is the sum of the propagation delays of the ADCMP582 comparator chip (U1) and MC10ELT21DG differential PECL to TTL translator chip (U2), 180 ps and 3.5 ns, respectively \cite{L-Datasheet-ADCMP582,M-Datasheet-MC10ELT21DG}. The measured digitizing time was (2.96$\pm$0.13) ns.

\section{Results from the experimental setup}
The Hamamatsu Multi-Pixel Photon Counter S12572-100P operation range frequency was measured to be from 1 Hz to 10 kHz, and to be from 1 Hz to 100 kHz, using a green LED light pulsed with a decay exponential waveform voltage and using a pulse waveform with a duty cycle of 1.5 \%, respectively. We found no dark counts or spurious signals when the controlled LED light is turn off.

Exposing to natural radiation the detection material attached to Hamamatsu Multi-Pixel Photon Counter S12572-100P, we got these results: The noise signal was $\pm$20 mV\textsubscript{Pk-Pk}; the analog signal is positive exponential decay, with an amplitude of $\sim$280 mV\textsubscript{Pk-Pk}, a rise time of $\sim$13 ns, and a fall time of $\sim$249.60 ns. 

\begin{figure}[ht!]
\centering
\includegraphics[width=0.70\textwidth,trim=0 0 0 0,clip]{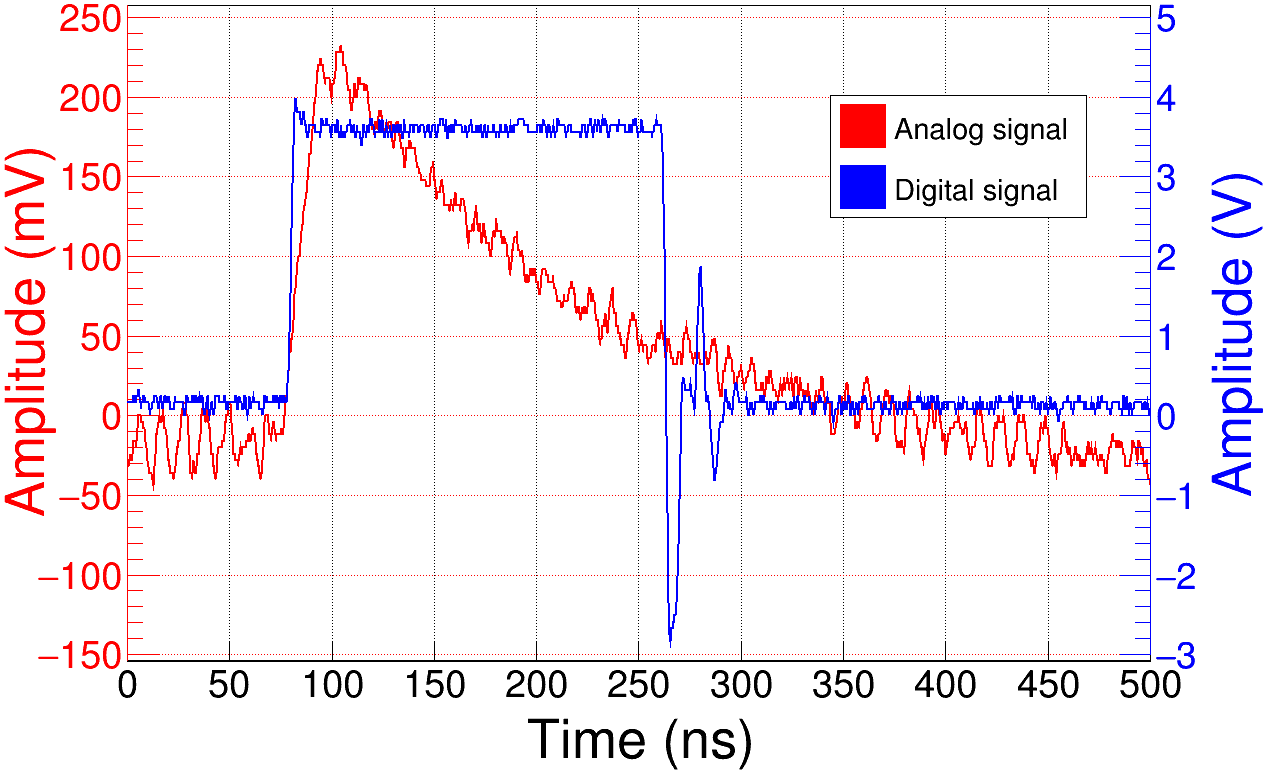}
	  \caption{\label{Fig:WaveScr60mV} Waveform from experimental setup using threshold voltage at $+$60 mVdc and feeding voltage at $+$75 Vdc; the analog signal is the red waveform; the digital signal, the blue waveform.}
\end{figure}

The output digital pulse signal has $\sim$3.8 V\textsubscript{Pk-Pk} of amplitude, which is close to the operating voltage of the LVTTL family. In addition, we detected a positive overshooting at $+$2 Vdc and negative overshooting at $-$3 Vdc, both less than $\sim$10 ns width. This overshoot does not affect the cRIO measurements because the positive overshoot is less than $+$3.3 Vdc (LVTTL family) and less than 25 ns of width (operation rate of cRIO); see the figure~\ref{Fig:WaveScr60mV}; The NI‑9402 C module (from the cRIO) discriminates the negative overshoot. 

\section{Conclusions}
We presented the design, construction, characterization, and testing of an one-channel electronic module for the Hamamatsu Multi-Pixel Photon Counter S12572-100P.

We have evaluated transmission coefficient, phase shift, and transit time of feeding and reading-out electronic board. The average transmission coefficient is 0.77$\pm$0.01. The phase shift is almost zero degrees and independent on the frequency, but were randomly produced at some frequencies; at frequencies near 7.90 kHz and 8.00 kHz, the phase shift is $+$1.13 degrees and $-$1.71 degrees, respectively; at frequencies near 370 kHz and 600 kHz, the phase shift is $+$1.184 degrees and $+$1.171 degrees, respectively; at frequencies between 780 kHz and 810 kHz, there are four values for the phase shift, the first two with $-$5.17 degrees and the rest with values of $-$5.11 degrees. The measured transit time is (741.72$\pm$82.80) ps.

We have evaluated digitizing efficiency error ($eff$), digitizing efficiency, and digitizing time of digitizing-electronic board. The average $eff$ is 0.88$\times10^{-3}$ \%. The digitizing efficiency is 99.99 \%. The digitizing time is (2.96$\pm$0.13) ns.

From the experimental setup used to evaluate the Hamamatsu Multi-Pixel Photon Counter S12572-100P, the digital signal corresponds perfectly to the analog signal; the characteristic results of the analog signal obtained by Hamamatsu Multi-Pixel Photon Counter S12572-100P and its electronic module are very similar to the characteristic results published by the Hamamatsu Multi-Pixel Photon Counter S12572-100P datasheet.

The transmission coefficient can be used to correct the analog output signal of the electronic board. The phase shift is low and negligible for this work. The transit time indicates a high-speed electronic module able to read high-speed signals and avoid the superposition of signals and errors when transmitted by other electronic circuit states.

The digitizing efficiency error is low and negligible for this work. The digitizing efficiency indicates that the design and manufacture of this electronic board are reliable. The digitizing time measured indicates discrimination or conversion of a very-narrow analog signal, and its value is lower than the one reported by manufacturers.

An interesting aspect of this electronic module is that any other MPPC series or model can be used with minor changes in the feeding and reading-out electronic board. The production cost is approximately 68 USD (not including the Hamamatsu Multi-Pixel Photon Counter S12572-100P and metal shielding costs).

We manufactured, tested, and run a set of ten feeding and reading-out electronic boards and five digitizing-electronic boards, the obtained results were similar.

The MPPC module developed is a robust, high-speed, and low-cost that can be used in high-energy physics applications like cosmic ray and neutrino detectors, dark matter, calorimeters, and Cherenkov telescopes, among others.

\bibliographystyle{JHEP}
\bibliography{refs}

\providecommand{\href}[2]{#2}\begingroup\raggedright\begin{thebibliography}{10}

\bibitem{A-Hamamatsu1}
Hamamatsu, ``Datasheet.''
  \url{https://www.hamamatsu.com/content/dam/hamamatsu-photonics/sites/documents/99_SALES_LIBRARY/ssd/mppc_kapd9008e.pdf}.

\bibitem{B-Gundacker_2020}
S.~Gundacker and A.~Heering, \emph{The silicon photomultiplier: fundamentals
  and applications of a modern solid-state photon detector},
  \href{https://doi.org/10.1088/1361-6560/ab7b2d}{\emph{Physics in Medicine
  Biology} {\bfseries 65} (2020) 17TR01}.

\bibitem{C-Hamamatsu2}
Hamamatsu, ``Datasheet.''
  \url{http://laboratoriodeparticulaselementales.blogspot.com/p/documentos.html}.

\bibitem{D-Hamamatsu3}
Hamamatsu, ``Si detectors for high energy particle.''
  \url{https://www.hamamatsu.com/content/dam/hamamatsu-photonics/sites/documents/99_SALES_LIBRARY/ssd/high_energy_kspd9002e.pdf},
  2021.

\bibitem{E-YOKOYAMA2009128}
M.~Yokoyama, T.~Nakaya, S.~Gomi, A.~Minamino, N.~Nagai, K.~Nitta et~al.,
  \emph{Application of hamamatsu mppcs to t2k neutrino detectors},
  \href{https://doi.org/https://doi.org/10.1016/j.nima.2009.05.077}{\emph{Nuclear
  Instruments and Methods in Physics Research Section A: Accelerators,
  Spectrometers, Detectors and Associated Equipment} {\bfseries 610} (2009)
  128}.

\bibitem{F-UserManual-cRIO-9025}
``Ni crio-9025 intelligent real-time embedded controller for compactrio.''
  \url{http://www.ni.com/pdf/manuals/375490d.pdf}, October, 2015.

\bibitem{G-UserManual-cRIO-9118}
``Ni crio-9111/9112/9113/9114/9116/9118 compactrio reconfigurable embedded
  chassis.'' \url{http://www.ni.com/pdf/manuals/375079e.pdf}, April, 2016.

\bibitem{H-Datasheet-cRIO-9402}
``Ni 9402, 4 dio, lvttl, bidirectional, 55 ns.''
  \url{http://www.ni.com/pdf/manuals/374614a_02.pdf}, January, 2016.

\bibitem{I-UserManual-TDS2022C-EDU}
``Tds2000c and tds1000c-edu series digital storage oscilloscopes user manual.''
  \url{https://download.tek.com/manual/TDS2000C-and-TDS1000C-EDU-Oscilloscope-User-Manual-EN.pdf},
  April 11, 2013.

\bibitem{J-use_of_an_oscilloscope}
D.J.~Mowbray, ``Use of an oscilloscope, webside.''
  \url{https://www.yumpu.com/en/document/view/18116200/use-of-an-oscilloscope},
  July, 2003.

\bibitem{K-coughlin2001operational}
R.~Coughlin and F.~Driscoll, \emph{Operational Amplifiers and Linear Integrated
  Circuits}, Prentice-Hall, Inc., Upper Saddle River, New Jersey, USA, 07458.
  (2001).

\bibitem{L-Datasheet-ADCMP582}
A.~Devices, ``Datasheet.''
  \url{http://www.analog.com/media/en/technical-documentation/data-sheets/ADCMP580_581_582.pdf},
  2016.

\bibitem{M-Datasheet-MC10ELT21DG}
O.~Semiconductor, ``Datasheet.''
  \url{https://www.mouser.com/datasheet/2/308/1/MC10ELT21_D-2315134.pdf},
  August, 2015.

\end{thebibliography}\endgroup

\end{document}